\documentclass[10pt]{article}

\usepackage[super]{cite}

\usepackage{sectsty}

\sectionfont{\fontsize{12}{14}\bfseries\sffamily}

\usepackage{amssymb}
\usepackage{amsfonts}
\usepackage{amsmath}

\usepackage{xcolor}

\definecolor{garrosgreen}{rgb}{0.1, 0.4, 0.1}
\definecolor{dartmouthgreen}{rgb}{0.05, 0.5, 0.06}
\definecolor{duelferred}{rgb}{0.7, 0.2, 0.1}
\definecolor{oxfordblue}{rgb}{0.05, 0.2, 0.7}

\usepackage{siunitx}

\newcommand{\calC}{\mathcal{C}}

\newcommand{\plus}{{\mbox{{\bf{\tiny +}}}}}

\newcommand{\rmT}{{\rm T}}

\usepackage{bbm}

\def\dd{{\mathrm{d}}}
\def\ee{{\mathrm{e}}}
\def\ii{{\mathrm{i}}}

\def\calO{\mathcal{O}}

\begin{document}

{\large \textbf{\textsf{Equivalence Principle for Antiparticles and Its Limitations}}}

{\rm U. D. Jentschura}

{\em \scriptsize Department of Physics,
Missouri University of Science and Technology,
Rolla, Missouri 65409, USA}

\begin{abstract}
We investigate the particle-antiparticle symmetry of the
gravitationally coupled Dirac equation,
both on the basis of the gravitational central-field
problem and in general curved space-time backgrounds.
First, we investigate the central-field
problem with the help of a Foldy--Wouthuysen
transformation. This disentangles the
particle from the antiparticle solutions,
and leads to a ``matching relation'' of the
inertial and the gravitational mass, which is
valid for both particles as well as antiparticles.
Second, we supplement this derivation
by a general investigation of the behavior of the
gravitationally coupled Dirac equation under
the discrete symmetry of charge conjugation,
which is tantamount to a particle$\to$antiparticle
transformation. Limitations of the Einstein equivalence
principle due to quantum fluctuations are discussed.
In quantum mechanics, the question of where and when in the Universe
an experiment is being performed,
can only be answered up to the limitations implied by Heisenberg's
Uncertainty Principle, questioning an assumption made
in the original formulation of the Einstein equivalence
principle. Furthermore, at some level
of accuracy, it becomes impossible to separate
non-gravitational from gravitational experiments,
leading to further limitations.
\end{abstract}


\small

\tableofcontents

\normalsize

%
%
\section{Introduction}
\label{sec1}

A significant
motivation for antimatter tests of gravity
is that the equivalence principle
has never been tested for antiparticles.
We aim to give an account of the implications of the gravitationally 
coupled Dirac equation for antiparticles,
motivated by a couple of ongoing
experiments at CERN~\cite{ALPHA,%
AmEtAl2011,AmEtAl2013,%
ATHENA,%
ATRAP,GaEtAl2008,%
Ke2008,%
AGELOIshort,AGELOI}.
Here, we investigate, in detail, the relation
of the inertial mass of spin-$1/2$ particles and 
antiparticles, based on the Dirac equation 
which, as is well known to atomic theorists~\cite{MoPlSo1998}
but perhaps less well known in the general relativity
community, describes both particles as well as the 
corresponding antiparticles simultaneously.
We observe that, {\em a priori}, it is the inertial, not the gravitational,
mass which enters the Dirac equation.
However, we will attempt to show that,
by coupling the Dirac equation to curved space-time, 
it is possible to identify the relation of the 
inertial mass to the gravitational mass,
for both particles and antiparticles.
A number of consequences of our results are discussed.

Dirac is sometimes quoted as saying 
that the equation named after him~\cite{Di1928a,Di1928b}
is ``more intelligent than its inventor''.
Indeed, the Dirac equation solved a number 
of problems simultaneously:
{\em (i)} It provided the necessary linearization
of the Klein--Gordon equation, thus solving problems connected
with the negative probabilities otherwise encountered 
in the context of the Klein--Gordon equation.
The linearization also enabled physicists to 
formulate a Hamiltonian operator
for spin-$1/2$ particles~\cite{ItZu1980}.
{\em (ii)} The Dirac equation immediately led to 
a consistent description of an intrinsic degree
of freedom of an elementary particle, namely, the 
spin of the electron
(and also, of the spin of the positron).
{\em (iii)} The Dirac equation 
predicted the existence of 
particles of the same (inertial!) mass,
but opposite charge, namely, 
the existence of spin-$1/2$ antiparticles.
Indeed, the existence of the positron 
was confirmed after the Dirac equation was
invented~\cite{An1933}.
For an illustrative and interpretive
discussion of some interesting properties of the
Dirac equation, see Ref.~\citen{Ma2016}.

One might ask: If the Dirac equation makes a definite
prediction about the {\em charge} of the antiparticles,
then why could the Dirac equation not be used in order 
to make a definitive prediction about the 
{\em gravitational mass} of the antiparticles?
One should remember that conceivable
gravitational matter-antimatter repulsion has been
investigated in a number of theoretical papers
(see, e.g., Ref.~\citen{Ko1996}).

The answer is as follows:
{\em (i)} 
First, one has to realize that the mass which 
enters the free Dirac equation (no coupling to either 
electromagnetic or gravitational fields), constitutes the 
inertial (not the gravitational) mass of the particle.
After a transformation which disentangles the 
particle from the antiparticle,
known as the Foldy--Wouthuysen transformation~\cite{FoWu1950},
one can derive the energy-momentum relation for 
both particles as well as antiparticles.
One then immediately realizes that the Dirac 
equation predicts the same {\em inertial} mass for 
both kinds of particles.

{\em (ii)} The second step is to couple the Dirac particle
to an external electromagnetic field, 
and carry out the Foldy--Wouthuysen transformation,
which in this case becomes a little more involved.
After disentangling the particle from the 
antiparticle Hamiltonian, one realizes that the electric charge 
of the Dirac antiparticle has to be equal in magnitude,
but opposite in sign, to that of the particle.
In order to address a possible misunderstanding 
right from the start, let us also remember that,
in order to derive antiparticle properties from the
Dirac equation, it is not necessary to quantize the 
Dirac field: Namely, the negative-energy solutions 
of the Dirac equation are interpreted as 
antiparticle solutions, in view of the 
reinterpretation principle~\cite{ItZu1980,JeWu2012epjc}.
In that sense, there is no such thing as a ``single-particle
Dirac theory'': Rather, the solutions of the 
equation itself describe
particles (positive energy) as well as 
antiparticles (negative energy).
Recall that, historically, the prediction of the 
existence of the positron (with all of its
electromagnetic as well as {\em inertial} properties) 
on the basis of the Dirac equation~\cite{Di1928a,Di1928b} 
was followed by the discovery of the positron~\cite{An1933},
long before the concept field-quantization was introduced
in theoretical physics~\cite{Be1947}.

{\em (iii)} 
The third step then is to couple the Dirac equation
to a gravitational field. Here, too, 
a possible misunderstanding needs to be addressed
right from the start: One might think that
it is necessary to quantize gravity. 
However, that is not the case. General relativity is a classical
theory which promotes space-time to a curved structure;
events are described by space-time coordinates.
In consequence, the wave function of a particle
(in general relativity) becomes a function of the 
space-time coordinates~\cite{Je2013,JeNo2013pra,JeNo2014jpa,%
NoJe2015tach,NoJe2016,Je2018geonium}.
One then has to couple the Dirac particle to the curved
space-time. The foundations of that endeavor have
been laid by Tetrode~\cite{Te1928}, Fock and Ivanenko~\cite{FoIw1928,Fo1929,Fo1929crasp},
Weyl~\cite{We1929}, and Brill and Wheeler~\cite{BrWh1957}
(for an excellent historic account of the early
developments, see Ref.~\citen{Bl2018}). 
One needs to introduce a covariant coupling of the Dirac
particles, with respect to changes in local 
Lorentz frames~\cite{Fo1929,Bo2011}.
These changes in local Lorentz frames (space-time is locally flat)
assume the role of a (local) $SO(1,3)$ gauge 
transformation, which can be formulated
for spin-$1/2$ particles~\cite{Fo1929,BrWh1957,Iv1969b,IvMiVl1985,Bo2011}.
This is analogous to
the electromagnetic case, where the gauge 
transformation is that of the underlying $U(1)$ gauge theory.
Expressed differently,
the gravitational field, in this case, 
assumes the same role as a classical, external,
electromagnetic field entering the Dirac equation
(e.g., an external Coulomb field),
and does not need to be quantized.

After applying the Foldy--Wouthuysen transformation
to the gravitationally coupled Dirac equation, 
one obtains the effective particle and antiparticle 
Hamiltonians. Expressed differently, one starts with a theory in which the 
mass parameter assumes the role of the 
inertial mass, for both particles as well as 
antiparticles, and hopes to end up with two Hamiltonians:
the first of these is applicable to the particle, 
while the second Hamiltonian describes the 
gravitational coupling of the antiparticle.
It is then of interest to analyze in which manner 
the (inertial) mass of the particle (and anti-particle)
enter the final, gravitationally coupled
effective particle and antiparticle Hamiltonians, 
i.e., to investigate the relationship of the gravitational and 
inertial masses. Under favorable circumstances, one 
should be able to determine the functional relationship of the inertial 
mass and the gravitational mass, 
for both the particle and the antiparticle.

Again, in order to avoid further possible misunderstandings,
we should remember that, quite recently, 
in very remarkable experiments, 
the equality of the charge-to-{\em{inertial}}-mass ratio
of protons and antiprotons has been experimentally
verified to unprecedented precision~\cite{GaEtAl1999,UlEtAl2014}.
These results, however, do not have any connection
to the {\em gravitational} mass of the antiparticles,
which is the subject of the current investigation.

As a last possible interjection, one might point out that 
many particles used in fundamental physics experiments
on the properties of antimatter, actually constitute
composite particles. E.g., the mass of anti-protons
mostly is the mass equivalent of the extra energy
of the quarks and gluons in a region within the anti-proton,
as compared to the rest energy of the quarks 
alone in the quantum chromodynamic vacuum, 
accounts for about 99\% of the mass of the proton.
As the free quarks contribute little
to the hadron masses, the identical inertial masses of protons and antiprotons
are thus due mostly to the color (and electroweak) 
interactions which have the same
strengths between charges as between anticharges
(proton versus anti-proton).
In the context of the current investigation, what
matters is that, for purposes of the description
of quantum electrodynamic bound states,
protons (and anti-protons) can be described
to excellent accuracy by an effective Dirac equation
which includes form factors (see, e.g., Ref.~\citen{Je2011radii}).
On the level of the effective Dirac 
equation, protons and anti-protons 
are thus amenable to an analysis based on the same
formalism as originally devised for point-like
spin-$1/2$ particles (electrons and positrons).

We organize the paper as follows.
In Sec.~\ref{sec2}, 
we discuss the derivation 
of the gravitationally coupled Dirac 
equation in great detail.
The equivalence principle for antimatter
is explored in Sec.~\ref{sec3}.
Conclusions are reserved for Sec.~\ref{sec5}.
Three appendices round off the paper.
Appendix~\ref{appa}~is devoted to a discussion
of the differences between the gravitational and 
electrostatic central-field problems.
Limitations of Einstein's Equivalence Principle (EEP)
due to quantum effects are discussed in Appendix~\ref{appb}.
The relationship of our investigations
to the Penrose conjecture is treated in Appendix~\ref{appc}.
Units with $\hbar = \epsilon_0 = c = 1$
are used throughout the paper unless otherwise stated.

%
%
\section{Formalism for Gravitational Coupling}
\label{sec2}

%
%
\subsection{Free Dirac Equation}
\label{sec21}

First of all, we should realize that the 
mass term in the free Dirac equation~\cite{Di1928a,Di1928b,ItZu1980},
which in covariant form reads as
\begin{equation}
\label{dirac_eq}
(\ii \gamma^\mu \partial_\mu - m_I) \psi = 0 \,.
\end{equation}
The mass parameter in the free Dirac equation
is the inertial mass $m_I$ (further remarks on this 
point are given in the following).
Throughout this article, we use the 
Dirac matrices $\gamma^\mu$ in the Dirac representation,
\begin{equation}
\label{dirac_rep}
\gamma^0 = \left( 
\begin{array}{cc}
\mathbbm{1}_{2 \times 2} & 0 \\
0 & -\mathbbm{1}_{2 \times 2} 
\end{array}
\right) \,,
\qquad
\vec\gamma = \left( 
\begin{array}{cc}
0 & \vec \sigma \\
-\vec \sigma & 0 
\end{array}
\right) \,,
\end{equation}
where the $\vec\sigma$ are the Pauli matrices.
We can convert this equation into a non-covariant form,
and write the free Dirac Hamiltonian $H_{\rm FD}$ as
\begin{equation}
\label{HFD}
H_{\rm FD} = \vec \alpha \cdot \vec p + \beta m_I \,,
\qquad
\beta = \gamma^0 \,,
\qquad
\vec\alpha = \beta \, \vec\gamma \,.
\end{equation}
Let us calculate the square
of the non-covariant form of the free Dirac equation
$\ii \partial_t \psi = H_{\rm FD} \psi$.
Using the fact that $\{ \vec\alpha, \beta\} = 0$,
one obtains
\begin{equation}
-\partial^2_t \psi = [(\vec \alpha \cdot \vec p + \beta m_I)^2] \psi 
= \left( \vec p^{\,2} + m_I^2 \right) \psi \,.
\end{equation}
For stationary states, one can replace 
$\ii \, \partial_t \to E$,
and thus $-\partial^2_t \to E^2$.
The dispersion relation
\begin{equation}
E = \pm \sqrt{\vec p^{\,2} + m_I^2 } = 
\pm \left( 
m_I + \frac{\vec p^{\,2}}{2 m_I} - \frac{\vec p^{\,4}}{8 m_I^3} 
+ \dots \right) 
\end{equation}
dictates that $m_I$ should be interpreted as the 
inertial mass, not the gravitational mass.
Restoring SI units temporarily, we see that the first
term $m_I$ is just the rest-mass energy $m_I \, c^2$, 
while the second term describes the Schr\"{o}dinger energy,
and the third gives the relativistic correction.

In fact, the (unitary) Foldy--Wouthuysen (FW) transformation~\cite{FoWu1950}
brings the free Dirac (FD) Hamiltonian into diagonal form,
\begin{align}
\label{UHFDU}
H^{\rm FW}_{\rm FD} =& \; U \, H_{\rm FD} \, U^{-1} = 
\left( \begin{array}{cc}
\sqrt{\vec p^{\,2} + m_I^2 } \, \mathbbm{1}_{2 \times 2} & 0 \\
0 & -\sqrt{\vec p^{\,2} + m_I^2 } \, \mathbbm{1}_{2 \times 2}
\end{array} \right) 
\nonumber\\[0.1133ex]
=& \; \left( \begin{array}{cc}
H_{\rm FD}^+ & 0 \\
0 & -H^-_{\rm FD}
\end{array} \right) \,,
\end{align}
where $\mathbbm{1}_{2 \times 2}$ is the two-dimensional unit matrix.
Here, $H_{\rm FD}^+$ is the $(2 \times 2)$ particle Hamiltonian,
while $H_{\rm FD}^-$ is the $(2 \times 2)$ anti-particle Hamiltonian.
The minus sign is due to the reinterpretation principle~\cite{BjDr1964,BjDr1965}
for anti-particles, which implies that an antiparticle
with eigenvalue $-E$ of the time derivative operator 
$\ii \partial_t$ is interpreted as an anti-particle
state with physical energy $E$.

The transformation $U$ is constructed, in the general case,
as
\begin{equation}
\label{defSandU}
U = \ee^{\ii S} \,,
\qquad
S = -\ii \, \beta \, \frac{\calO}{2 m_I} \,,
\end{equation}
where $\calO$ is the ``odd'' part [in $(2 \times 2)$ bispinor space]
of the Dirac Hamiltonian.
In general, the elimination of the odd operators
is based on the identity
\begin{equation}
[\beta, \beta \calO] = 2 \, \calO \,,
\end{equation}
which holds for a general odd matrix $\calO$.
For the free Dirac Hamiltonian~\eqref{HFD}, one sets
\begin{equation}
\calO = \vec\alpha \cdot \vec p \,,
\end{equation}
which eliminates the term of first order in the 
momenta in Eq.~\eqref{HFD}.
The relativistic correction terms are obtained if one uses 
instead
\begin{align}
\calO =& \; \vec\alpha \cdot \vec p \, \theta(|\vec p|) \approx 
\vec\alpha \cdot \vec p -
\frac{(\vec\alpha \cdot \vec p)^3}{ 6 m_I^2 } \,,
\nonumber\\[0.1133ex]
\theta(|\vec p|) =& \;
\frac{1}{2 | \vec p |} \,
\arctan\left( \frac{ | \vec p | }{m_I} \right) \,.
\end{align}
In summary, one obtains
the equivalent free Dirac Hamiltonians
$H_{\rm FD}^+$ and $H_{\rm FD}^-$ for 
particles and antiparticles,
\begin{equation}
H_{\rm FD}^+ = H_{\rm FD}^- =
\sqrt{\vec p^{\,2} + m_I^2 } \, \mathbbm{1}_{2 \times 2} \,.
\end{equation}
The same inertial mass $m_I$ enters both 
$H_{\rm FD}^+$ and $H_{\rm FD}^-$.

%
%
\subsection{Covariant Gravitational Coupling}
\label{sec22}

The theoretical formulation of the 
gravitational coupling of Dirac particles was
developed when scientists were trying to 
understand the connection of Einstein's 
general theory of relativity to 
quantum mechanics~\cite{Te1928,FoIw1928,Fo1929,Fo1929crasp,%
We1929,BrWh1957} (see also Ref.~\citen{Bl2018});
a particularly pointed formulation
is the ``geometrization'' 
of ``Dirac's theory of the electron''
as the title of Ref.~\citen{Fo1929} suggests.
The positron was not mentioned in the 
title of the publication~\cite{Fo1929},
for obvious reasons:
it had not even been discovered yet~\cite{An1933}.

Now, let us try to present a derivation of 
the gravitational coupling terms for 
antiparticles; we will concentrate on the leading terms,
namely, those relevant to the non-relativistic 
limit. The gravitational coupling of the Dirac equation 
entails two replacements as compared to 
Eq.~\eqref{dirac_eq},
\begin{equation}
\label{repl}
\gamma^\mu \to {\overline \gamma}^\mu \,,
\qquad
\partial_\mu \to
\nabla_\mu = \partial_\mu - \Gamma_\mu \,,
\end{equation}
where the 
curved-space Dirac matrices ${\overline \gamma}^\mu$
fulfill ``local'' commutation relations given in
Eq.~\eqref{local}. These depend
on the space-time coordinates. The spin-connection 
matrix $\Gamma_\mu$ describes the space-time curvature.
The Dirac equation assumes the form 
\begin{equation}
(\ii {\overline \gamma}^\mu \nabla_\mu - m_I) \psi = 0 \,.
\end{equation}
under gravitational coupling.
We recall that the free-space Dirac matrices fulfill
\begin{equation}
\{ \gamma^\mu, \gamma^\nu \} = 2 g^{\mu\nu} = 
2 \, {\rm diag}(1,-1,-1,-1) \,.
\end{equation}
By contrast, the curved-space Dirac matrices fulfill
\begin{equation}
\label{local}
\{ {\overline \gamma}^\mu, {\overline \gamma}^\nu \} = 2 \,
{\overline g}^{\mu\nu}(x) 
\end{equation}
where ${\overline g}^{\mu\nu}(x)$ is the metric 
of curved space-time. One sets~\cite{Fo1929,BrWh1957,Iv1969b,IvMiVl1985}
\begin{equation}
{\overline \gamma}^\mu = e^\mu_A \, \gamma^A \,,
\end{equation}
where the $e^\mu_A$ are the coefficients
of the {\em vierbein} or ``tetrad''~\cite{Iv1969b,IvMiVl1985}, 
and coefficients
in the anholonomic basis are denoted by capital Latin
letters $A, B, \ldots = 0,1,2,3$.

Roughly speaking, the derivation of the 
gravitational coupling proceeds as follows.
As space-time is locally flat,
one should be able to change the local Lorentz frames, and 
therefore, the {\em vierbein} coefficients,
without changing the physics. By comparison,
for quantum electrodynamics~\cite{ItZu1980}, one is able to change the
local phase of the wave function independently 
at any point in space-time, provided one also 
performs a concomitant gauge transformation of the 
electromagnetic four-vector potentials.

On the basis of these considerations, one
can derive a spinor representation
of the local Lorentz group.
One then formulates a covariant derivative
with respect to the group $SO(1,3)$ of local 
Lorentz transformation, 
in accordance with the gauge principle. The paradigm is that 
the covariant derivative of the 
spinor-Lorentz transformation of the Dirac bispinor,
formulated in the transformed coordinates,
is equal to the spinor-local-Lorentz-transformation
of the covariant derivative of the same Dirac bispinor.
This can be formulated as follows,
\begin{equation}
\label{nablaprime}
\nabla'_\mu \psi' = 
(\nabla_\mu \psi)' \,,
\qquad
\psi' = S(\Lambda) \, \psi \,,
\qquad
(\nabla_\mu \psi)' =
S(\Lambda) \, \nabla_\mu \psi \,.
\end{equation}
A closer inspection~\cite{Bo2011,JeAd2019} of the 
problem reveals that this condition 
is fulfilled if one sets
\begin{equation}
\label{Gammamu_def}
\Gamma_\mu = \frac{\ii}{4} \, \omega_\mu^{AB} \, \sigma_{AB} \,,
\qquad
\omega^{AB}_\mu = e^A_\mu \nabla_\mu e^{\nu B} \,,
\end{equation}
where the Ricci rotation coefficients are 
denoted as $\omega^{AB}_\mu$, and the 
spin matrices are $\sigma_{AB} = \frac{\ii}{2} \, 
[ \gamma_A, \gamma_B ]$. The covariant 
derivative acts as follows,
\begin{equation}
\nabla_\mu e^{\nu B} =
\partial_\mu e^{\nu B} + \Gamma^\nu_{\mu\rho} e^{\rho B} \,.
\end{equation}
Here, the $ \Gamma^\nu_{\mu\rho}$ are the Christoffel symbols.

A change in the vierbein $e^\mu_A \to {e'}^\mu_A$
amounts to a local Lorentz transformation,
and the $\nabla'_\mu$ in Eq.~\eqref{nablaprime} 
is understood as $\nabla'_\mu = \partial_\mu - \Gamma'_\mu$ 
(which entails no change in $\partial_\mu$, 
but a change in the Ricci rotation coefficients),
\begin{equation}
\Gamma'_\mu = \frac{\ii}{4} \, {\omega'}_\mu^{AB} \, \sigma_{AB} \,,
\qquad
{\omega'}^{AB}_\mu = {e'}^A_\mu \nabla_\mu {e'}^{\nu B} \,.
\end{equation}

%
%
\section{Equivalence Principle for Antiparticles}
\label{sec3}

%
%
\subsection{Central--Field Problem}
\label{sec31}

The general formalism outlined in Sec.~\ref{sec22} has to be applied 
in practice. In a ``weak'' gravitational field 
described by a potential $\Phi$, one has the
metric~\cite{Wi1974prd,Je2018geonium},
\begin{align}
\label{metric}
\dd s^2 =& \; g_{\mu\nu} \dd x^\mu \, \dd x^\nu
= \left( 1 + 2 \, \Phi \right) \, \dd t^2
- \left( 1 - 2 \, \Phi \right) \, \dd \vec r^{\,2} 
\nonumber\\[0.1133ex]
=& \; \left( 1 - \frac{ 2 \, G M}{r} \right) \, \dd t^2
- \left( 1 + \frac{ 2 \, G M}{r} \right) \, \dd \vec r^{\,2} \,,
\end{align}
where in the case of a central gravitational potential,
one has $\Phi = -G M/r$.
An evaluation of the Christoffel symbols,
and of the spin connection matrices, leads to the 
following result~\cite{Wi1974prd,JeNo2013pra,Je2018geonium},
which holds to first order in the gravitational
coupling constant,
\begin{align}
\label{HG}
H_G =& \;
\frac12 \, \left\{ 1 + 2 \Phi,
\vec\alpha \cdot \vec p \right\} 
+ \beta m_I \left( 1 + \Phi \right) 
\\[0.1133ex]
=& \; \frac12 \, \left\{ 1 - \frac{ 2 \, G m M}{r}, 
\vec\alpha \cdot \vec p \right\} 
+ \beta m_I \left( 1 - \frac{G M}{r}  \right) \,.
\nonumber
\end{align}

At the risk of oversimplification,
we can say that the anti-commutator term 
involving $\vec\alpha \cdot \vec p$ comes from the 
gravitational covariant derivative, 
while the term 
$\beta m \left( 1 - G m M/r  \right)$ 
is due to the replacement of the 
flat-space Dirac matrices by their 
curved-space equivalents.

The functional form of the second term in 
Eq.~\eqref{HG} can be derived very easily 
by considering the 
Dirac equation for particles at rest,
where there is no momentum operator at all.
One applies the replacement~\eqref{repl}
to the Dirac matrices,
in the form $\gamma^0 \to {\overline \gamma}^0$. 
The free Dirac equation for particles and antiparticles
at rest is simply
\begin{equation}
\ii \gamma^0 \partial_t \psi = m_I \, \psi  \,.
\end{equation}
In curved space (central-field problem), one replaces
\begin{equation}
\ii \gamma^0 \partial_t \psi = m_I \, \psi \;\; \to \;\;
\ii {\overline \gamma}^0 \, \partial_t \psi = m_I \, \psi \,,
\end{equation}
where in view of Eq.~\eqref{metric},
\begin{equation}
\label{gravprep}
{\overline \gamma}^0 = 
\sqrt{ {\overline g}^{00} } \, \gamma^0 =
\sqrt{ \frac{1}{1 + 2 \, \Phi } } \gamma^0 \approx
\left( 1 - \Phi \right) \gamma^0  \,. 
\end{equation}
In the latter term, we have expanded the square root to first 
order in $\Phi$.
One notes that ${\overline g}^{00}$ is 
a coefficient of the inverse metric ${\overline g}^{\mu\nu}$, 
where ${\overline g}^{\mu\nu} \, {\overline g}_{\nu\rho} = 
{\delta^\mu}_\rho$.
Hence, with ${\overline \gamma}_0 \, {\overline \gamma}^0 =  1$,
keeping track of upper and lower indices carefully, we have
\begin{equation}
\label{gravdirect}
\ii \partial_t \psi 
\approx m_I \, {\overline \gamma}_0 \, \psi
\approx \gamma^0 \, m_I \, (1 + \Phi) \psi 
= \beta \, m_I \, \left(1 - \frac{G M}{r} \right) \psi \,.
\end{equation}
{\em This simple consideration,
at the risk of some over-simplification,
rederives the second term on the right-hand side
of Eq.~\eqref{HG}.
It means that the leading gravitational interaction
term in a central field follows from 
the metric-induced modification of the 
Dirac $\gamma^0$ matrix alone,
without any recourse to Christoffel symbols
or gravitational connection matrices.}

In order to derive the equivalence principle
in leading order, it is sufficient to 
approximate Eq.~\eqref{HG} by
\begin{align}
\label{HGapprox}
H_G =& \;
\vec\alpha \cdot \vec p 
+ \beta m_I \left( 1 + \Phi \right) 
\nonumber\\[0.1133ex]
=& \; 
\vec\alpha \cdot \vec p 
+ \beta m_I \left( 1 - \frac{G M}{r}  \right)  \,,
\end{align}
because the kinetic term already contains a momentum
and is of higher order.

If we add a nontrivial potential to the
free Dirac Hamiltonian, as in Eq.~\eqref{HGapprox},
then it is not possible any more to
diagonalize the Dirac Hamiltonian in bispinor space
exactly. Rather, one employs a perturbative
approach, with a Foldy--Wouthuysen transformation that
eliminates the odd contributions,
order by order in the momenta~\cite{BjDr1964,ItZu1980}.
A single step of the FW transformation, using
[see Eq.~\eqref{defSandU}]
\begin{equation}
U = \exp\left( \ii S \right) \,,
\qquad
S = -\ii \beta \, \frac{\vec\alpha \cdot \vec p}{2 m_I} \,,
\end{equation}
then is sufficient to obtain
the Foldy-Wouthuysen transformed, gravitationally
coupled Hamiltonian
\begin{align}
H^{\rm FW}_G =& \; U \, H_G \, U^{-1} 
\nonumber\\[0.1133ex]
=& \; H_G + \ii [S, \, H_G ] +
\frac{\ii^2}{2!} \, [S, [S, \, H_G]] + \dots,
\end{align}
which reads as follows,
\begin{align}
H^{\rm FW}_G =& \;
\beta \left[ \frac{\vec p^{\,2}}{2 m_I} + m_I \, \Phi \right]
\nonumber\\[0.1133ex]
=& \; \left( \begin{array}{cc}
\displaystyle 
\left( \frac{\vec p^{\,2}}{2 m_I} - G \frac{m_I M}{r} \right) \,
\mathbbm{1}_{2 \times 2} & 0 \\
0 & 
\displaystyle 
-\left( \frac{\vec p^{\,2}}{2 m_I} - G \frac{m_I M}{r} \right) \,
\mathbbm{1}_{2 \times 2} 
\end{array}
\right)
= \left( \begin{array}{cc}
H^+_G & 0 \\
0 & -H^-_G 
\end{array}
\right).
\end{align}
So, the particle and antiparticle Hamiltonians are equal,
\begin{equation}
\label{42}
H^+_G = H^-_G = 
\left( \frac{\vec p^{\,2}}{2 m_I} - G \frac{m_I M}{r} \right) \,
\mathbbm{1}_{2 \times 2} \,.
\end{equation}
Note that the Newtonian central-field term $(- G m_I M/r)$
is obtained naturally in this Hamiltonian, for both 
particles as well as antiparticles.
The emergence of the matrix $\mathbbm{1}_{2 \times 2}$ implies that the 
energy is independent of the spin of the particle.

Let us remember that Newton stated that the property of a body called
``mass'' (``inertial mass'') has to be
proportional to the ``weight'' (which enters
the gravitational force law), a principle otherwise known
as the ``weak equivalence principle'' (WEP).
Upon choosing the physical value of the 
gravitational constant $G$ appropriately,
one finds the exact equivalence $m_G = m_I$
of the gravitational mass $m_G$ and the 
inertial mass $m_I$ of a particle.
The Einstein equivalence principle (EEP) states that {\em (i)} WEP is valid,
{\em (ii)} the outcome of any local non-gravitational experiment is independent
of the velocity of the freely-falling reference frame in which it is performed
(local Lorentz invariance, LLI), and
{\em (iii)} the outcome of any local non-gravitational experiment is
independent of where and when in the universe it is performed
(local position invariance, LPI).
 
We know that the classical Hamiltonian for a
particle subject to a central gravitational potential
is~\cite{Go1963}
\begin{equation}
\label{HGcl}
H_{G,{\rm cl}} = \left( \frac{\vec p^{\,2}}{2 m_I} 
- G \, \frac{m_G M}{r} \right) \,,
\end{equation}
where we neglect the spin (which does not exist in a classical theory),
and denote the gravitational mass by $m_G$.
By the correspondence principle,
the quantum analogue is obtained from Eq.~\eqref{HGcl}
by interpreting the momentum $\vec p$
in Eq.~\eqref{HGcl} as the momentum operator
$\vec p = -\ii \vec\nabla$.

A comparison of Eqs.~\eqref{42} and~\eqref{HGcl} reveals that
\begin{subequations}
\begin{align}
m_G =& \; m_I \qquad \mbox{(particles)} \,,
\\[0.1133ex]
m_G =& \; m_I \qquad \mbox{(anti-particles)} \,,
\end{align}
\end{subequations}
thus establishing the equivalence principle
in the Newtonian form, for antiparticles,
at least within a leading-order calculation.

%
%
\subsection{Coupling to a General Background}
\label{sec32}

We now investigate the general case, where the 
Dirac particle is not necessarily coupled to a 
static central field, but is coupled 
to a general (dynamic) background, 
with a (possibly time-dependent) 
space-time metric, which gives rise to 
time-dependent connection matrices $\Gamma_\mu$
[see Eq.~\eqref{Gammamu_def}].

From nonrelativistic quantum mechanics, we 
know that the electromagnetic coupling 
can be described by the covariant coupling 
$\vec p \to \vec p - e \, \vec A$,
where $\vec A$ is the vector potential of the 
electromagnetic field.
The general electromagnetically and gravitationally 
coupled Dirac equation, for arbitrary electromagnetic 
four-vector potentials $A_\mu$, and 
arbitrary connection matrices $\Gamma_\mu$, 
reads as
\begin{equation}
\label{ord}
\left[ 
\overline\gamma^\mu \left\{ \ii \left( 
\partial_\mu - \Gamma_\mu \right) - e \, A_\mu \right\} - 
m_I \right] \, \psi = 0 \,,
\end{equation}
where the four-vector potential has components $A^\mu = (\Phi, \vec A)$,
and $e$ denotes the electron charge.
We note the scalar potential $\Phi$
and the vector potential $\vec A$ (see Chap.~8 of Ref.~\citen{Je2017book}).
Recall the $\Gamma_\mu$ matrices from Eq.~\eqref{Gammamu_def}:
\begin{equation}
\Gamma_\mu = \frac{\ii}{4} \, \omega_\mu^{AB} \, \sigma_{AB} \,,
\qquad
\omega^{AB}_\mu = e^A_\mu \nabla_\mu e^{\nu B} \,.
\qquad
\sigma_{AB} = 
\frac{\ii}{2} \, [ \gamma_A, \gamma_B ] \,.
\end{equation}

Now, let us carry out the charge conjugation,
which is tantamount to the particle$\to$antiparticle 
transformation. 
Under transposition and complex conjugation, one obtains
\begin{equation}
\label{previous}
\psi^\plus \left( \left( 
\overline\gamma^\mu \right)^\plus
\left( -\ii \overleftarrow{\partial}_\mu - e \, A_\mu \right) 
+ \ii \,  \Gamma^\plus_\mu \,  \left( 
\overline\gamma^\mu \right)^\plus
- m_I \right) = 0 \,,
\end{equation}
where the differential operator acts on 
$\psi^\plus$, i.e., to the left.
We remember that $\Gamma_\mu$ is matrix-valued.
The Hermitian adjoint $\Gamma^\plus_\mu$ of $\Gamma_\mu$ is calculated
as follows,
\begin{equation}
\label{signchange}
\Gamma^\plus_\mu 
= -\frac{\ii}{4} \, \omega_\mu^{AB} \, \sigma^\plus_{AB} 
= -\frac{\ii}{4} \, \omega_\mu^{AB} \, \gamma^0 \, \sigma_{AB} \, \gamma^0 
= - \gamma^0 \, \Gamma_\mu \gamma^0 \,,
\end{equation}
where one has used the result 
\begin{align}
\sigma^\plus_{AB} 
=& \; -\frac{\ii}{2} \, [ \gamma^\plus_B, \gamma^\plus_A ] 
= -\frac{\ii}{2} \, \gamma^0  \;
[ \gamma^0 \gamma^\plus_B \gamma^0, \;
\gamma^0 \gamma^\plus_A \gamma^0 ] \; \gamma^0 
\nonumber\\[0.1133ex]
=& \; -\frac{\ii}{2} \, \gamma^0 \;
[ \gamma_B, \; \gamma_A ] \; \gamma^0 
= - \gamma^0 \; \sigma_{BA} \; \gamma^0 
= \gamma^0 \; \sigma_{AB} \; \gamma^0 \, .
\end{align}
Here, we recall that we use the Dirac representation, 
given in  Eq.~\eqref{dirac_rep}, where one has
the relationships
$\gamma^0 \left( \gamma^A \right)^\plus \gamma^0 =  \gamma^\mu $
and
$\gamma^0 \left( \overline\gamma^\mu \right)^\plus \gamma^0 =  \gamma^\mu $.
An insertion of $\gamma^0$ matrices, using the identity $(\gamma^0)^2 = 1$,
and multiplication by $\gamma^0$ from the right,
leads to
\begin{equation}
\left(  \psi^\plus \gamma^0 \right)
\gamma^0 \biggl[ \left( 
\overline\gamma^\mu \right)^\plus
\bigl( -\ii \overleftarrow{\partial}_\mu - e \, A_\mu \bigr)  
- \ii \gamma^0 \, 
\Gamma_\mu \gamma^0 \,  \left( 
\overline\gamma^\mu \right)^\plus 
- m_I \biggr]  \gamma^0 = 0 \,.
\end{equation}
For absolute clarity, observe that 
the sign change in the prefactor multiplying the 
$ \Gamma_\mu $ matrix, as compared to 
Eq.~\eqref{previous}, is due to Eq.~\eqref{signchange}.
This leads to the equation
\begin{equation}
\bar\psi 
\left[ 
\overline\gamma^\mu \left( -\ii \overleftarrow{\partial}_\mu 
- e \, A_\mu \right)
- \ii \, \Gamma_\mu 
\overline\gamma^\mu 
- m_I \right] = 0 \,.
\end{equation}
Under an additional transposition, one obtains
\begin{equation}
\label{sanse}
\left[ \left( 
\overline\gamma^\mu \right)^{\rmT} 
\left( -\ii \partial_\mu - e \, A_\mu \right) 
- \ii \left( 
\overline\gamma^\mu \right)^{\rmT} \Gamma^{\rmT}_\mu 
- m_I \right]
\bar\psi^{\rmT} = 0 \,.
\end{equation}
One introduces the charge conjugation matrix 
$C = \ii \, \gamma^2 \, \gamma^0$
(in the Dirac representation) with the 
properties
\begin{equation}
\label{DefCMatrix}
C  \left( \gamma^A \right)^{\rmT} C^{-1} = -\gamma^A \,,
\qquad
C  \left( \overline\gamma^\mu \right)^{\rmT} C^{-1} = -\overline\gamma^\mu \,,
\end{equation}
so that the charge conjugated 
spinor is $\psi^\calC(x) = C \, \bar\psi^{\rmT}(x)$.
In the Dirac representation, a possible choice is
$C = \ii \, \gamma^2 \, \gamma^0$.
Insertion of the identity matrix in the 
form $\mathbbm{1} = C^{-1} \, C$ in Eq.~\eqref{sanse} leads to
\begin{equation}
\left[ C \left( 
\overline\gamma^\mu \right)^{\rmT} C^{-1}
\left( -\ii \partial_\mu - e \, A_\mu \right) 
- \ii C \left( 
\overline\gamma^\mu \right)^{\rmT} C^{-1} \, 
C \, \Gamma^{\rmT}_\mu \, C^{-1} 
- m_I \right] \, C \, \bar\psi^{\rmT} = 0 \,.
\end{equation}
One calculates the identity
\begin{equation}
C \, \Gamma^{\rmT}_\mu \, C^{-1} = 
\frac{\ii}{4} \left\{ 
\frac{\ii}{2} \omega^{A B}_\mu \, 
C \, \left[ \gamma_B^{\rmT},\, \gamma_A^{\rmT} \right] \, C^{-1}
\right\}
= 
\frac{\ii}{4} \left\{ 
\frac{\ii}{2} \omega^{A B}_\mu \, 
\left[ -\gamma_B,\, -\gamma_A \right] 
\right\}
= -\Gamma_\mu \,.
\end{equation}
The charge conjugated 
(particle$\to$antiparticle transformed) Dirac equation fulfilled 
by $\psi^\calC(x)$ thus reads as 
\begin{equation}
\left[ \left(-
\overline\gamma^\mu \right)
\left( -\ii \partial_\mu 
- e \, A_\mu \right) 
- \ii \, 
\overline\gamma^\mu \, \Gamma_\mu - m_I \right] \;
\psi^\calC(x) = 0 \,,
\end{equation}
or alternatively as
\begin{equation}
\label{Ctransformed}
\left[ 
\overline\gamma^\mu 
\left\{ \ii (\partial_\mu - \Gamma_\mu) + e \, A_\mu \right\} 
- m_I \right] \; \psi^\calC(x) = 0 \,.
\end{equation}
The only difference between the original Dirac equation~\eqref{ord}
and Eq.~\eqref{Ctransformed} is the
sign of the physical charge $e$ of the particle.
The sign and the functional form of the gravitational coupling term
remains the same.

Note that $\psi^\calC(x) 
= C \, \bar\psi^{\rmT}(x)
= C \, (\psi^\plus \gamma^0)^{\rmT}(x)
= C \, \gamma^0 \psi^*(x)$
involves a complex conjugation. By way of example,
let us assume that $\psi(x)$ is proportional 
to $\exp(-\ii E t + \ii \vec k \cdot \vec r)$,
with a positive energy eigenvalue $E>0$.
In this case, $\psi^*(x)$ and, thus, $\psi^\calC(x)$
will be proportional to $\exp(\ii E t - \ii \vec k \cdot \vec r)$,
with a negative eigenvalue of the 
time derivative operator $\ii \partial_t$.
The latter form is the characteristic dependence for an 
antiparticle wave function.
Let us note that the comparison of Eqs.~\eqref{ord} and~\eqref{Ctransformed}
reveals that the gravitational coupling term 
is invariant under the particle$\to$antiparticle 
transformation in a general dynamic space-time 
background whose metric gives rise to Ricci 
rotation coefficients $\omega^{AB}_\mu$ and 
connection matrices $\Gamma_\mu$.

Furthermore, we
note that both Eqs.~\eqref{ord} and~\eqref{Ctransformed}
contain the inertial mass term $m_I$.
In principle, neither of these equations
say anything about the identification
of the inertial with the gravitational mass.
However, their comparison and the retention of the 
sign as well as the entire functional form of the 
gravitational coupling term in unchanged form
under the particle$\to$antiparticle transformation,
shows that if the equivalence $m_I = m_G$ 
for antiparticles
holds in {\em one} special example case (e.g.,
the central-field problem, see Sec.~\ref{sec22}), then it must hold for 
{\em any} general space-time geometry
(including all relativistic correction terms).
Additional considerations 
regarding the relativistic terms, for the special case
of a central-field problem, can be found in Ref.~\citen{Je2013}.

%
%
\section{Conclusions}
\label{sec5}

Let us briefly review the most important 
results of the current investigation.
In Sec.~\ref{sec2}, we present the formalism 
of gravitational coupling for spin-$1/2$ particles
on which this paper is based.
First, 
we recall that the Dirac Hamiltonian
describes particles and antiparticles
simultaneously (Sec.~\ref{sec21})
and recover a few details on the free Dirac equation.
The covariant coupling to curved space-time 
(gravitational coupling) is 
discussed in Sec.~\ref{sec22}.
We then derive 
the equivalence principle for antiparticles in an
illustrative way (in Sec.~\ref{sec3}), based on the Foldy--Wouthuysen
transformed Dirac Hamiltonian.
In Sec.~\ref{sec31},
the central-field problem is investigated,
and the derivation culminates in Eq.~\eqref{42},
where the particle and antiparticle
Hamiltonians in a 
central gravitational field are given.
Illustrative remarks on the 
differences between the electrostatic and 
gravitational central-field problems are given in Appendix~\ref{appa}.
The derivation is generalized to 
arbitrary curved space-time backgrounds
in Sec.~\ref{sec32},
for a general Dirac particle, coupled to 
both electromagnetic as well as gravitational
fields. The electromagnetic coupling term
reverses sign under the particle$\to$antiparticle
transformation, as it should, while the 
gravitational term retains its sign and 
its full functional form.
The conclusions of Sec.~\ref{sec22}
imply that $m_I = m_G$ for the central-field
problem (``matching''), and the considerations
of Sec.~\ref{sec32} generalize this finding to 
arbitrary (possibly dynamic) curved space times.

In summary, Sec.~\ref{sec3} is devoted to a 
derivation of the (weak) equivalence principle for antiparticles,
showing the equivalence of the inertial and gravitational
mass $m_I = m_G$ for antiparticles.
Within the Dirac formalism, any deviation of the gravitational mass
of antiparticles as compared to the gravitational 
mass of particles is thus constrained
by the very precise available comparisons
of the corresponding inertial masses,
which are available for some elementary 
particles~\cite{GaEtAl1999,UlEtAl2014}.
Conversely, if a deviation 
of the gravitational mass of an antiparticle
from that of a particle should be 
found in an upcoming experiment, then we would have
a clear motivation for fundamentally changing
our physical picture of the gravitational interaction
in general---not
only for antiparticles, but conceivably, also for particles.


This paper is rounded off by two more Appendices.
The first of these,
Appendix~\ref{appb}, is devoted to an
illustration of the limitations of the 
Einstein equivalence principle,
for both particles as well as antiparticles.
As already mentioned, the 
Einstein equivalence principle states, among 
other postulates, that 
the outcome of any local non-gravitational experiment is
independent of where and when in the universe it is performed
(local position invariance, LPI).
One of the ``compatibility issues'' of this postulate,
with respect to the uncertainty principle,
is that, at some level of precision, one cannot 
tell any more with absolute certainty where precisely the 
experiment was performed, because of the 
nonvanishing positional uncertainty of the 
wave function. Another ``incompatibility'' of the 
separation of non-gravitational and gravitational
experiments is found when we consider that
at some level of accuracy, one cannot separate a
non-gravitational from a gravitational experiment any more.
Even the electron spin, which would flip in a 
transition from, say, an $nP_{1/2}$ to an $nP_{3/2}$ 
level in hydrogen, couples to gravity 
(here, $n$ is the principal quantum number).
These observations lead to tiny deviation of
atomic clock comparisons at different places in 
a gravitational potential, from the formula~\eqref{EEP} 
that would otherwise be implied if the 
equivalence principle had no incompatibilities 
with quantum mechanics. Reassuringly, we can 
say that deviations from Eq.~\eqref{EEP},
given in Eq.~\eqref{corr_term},
are numerically small~\cite{Je2018geonium}.

Finally, we also look at the Penrose conjecture
in Appendix~\ref{appc}, which is motivated
by the observation that an ``uncertain'' quantum distribution
of mass, as implied by the Heisenberg uncertainty 
relation, implies an uncertainty in the solution of the 
Einstein equations due to the quantum effects.
The latter leads to an ``uncertainty'' in the 
determination of space-time curvature in the 
vicinity of the quantum mechanical wave function.
This uncertainty can only be resolved 
once the wave function collapses.
Our analysis, presented in Appendix~\ref{appc}, 
allows us to interpret the Penrose conjecture as the
``principle of reasonable determinability
of the space-time metric, in view of 
the uncertainty of the mass distribution associated with any 
quantum mechanical wave packet''.
We find that the collapse times 
implied by the Penrose conjecture
[see Eqs.~\eqref{tC} and~\eqref{EG1}],
for typical quantum systems, are longer than the 
age of the Universe. The good news derived 
from this observation is that,
apparently, in practical applications, 
the collapse of the wave function, 
due to other physical phenomena,
happens on such short time
scales that the concomitant uncertainty 
in the determination of the space-time curvature 
around the massive, quantum mechanical objects, 
is negligible.  If collapse always happens
on time scales shorter than would be 
postulated in the framework of the Penrose conjecture,
then the ``principle of reasonable determinability
of the space-time metric'' is always fulfilled
in practice---because of reasons that have
nothing to do with gravity but with
our apparent inability to
produce coherent superpositions of
macroscopic quantum objects.

\section*{Acknowledgments}

This work was supported by the 
National Science Foundation (Grant PHY--1710856). 
The author thanks G.~Gabrielse, A.~Geraci, T.~Kovachy, M.~Cavaglia, 
I.~N\'{a}ndori, J.~.H.~Noble and M.~Bojowald for useful discussions.
Part of this work was completed during a visit to 
the Center for Fundamental Physics at Northwestern 
University in Chicago; the author is grateful 
to G.~Gabrielse for the hospitality.

\appendix

\def\theequation{\thesection.\arabic{equation}}

\setcounter{equation}{0}

%
%
\section{(Na\"{\i}ve) Insertion of a Gravitational Potential}
\label{appa}

Let us briefly review why it is not consistent 
to simply insert the gravitational potential
into the Dirac Hamiltonian. This procedure 
would otherwise conceivably lead to a Hamiltonian 
of the functional form
\begin{equation}
\label{Hseriously}
\vec \alpha \cdot \vec p + \beta m_I 
- G \frac{m_G \, M}{r} \,,
\end{equation}
which differs from Eq.~\eqref{HGapprox}.
One might be tempted to consider the 
Hamiltonian~\eqref{Hseriously}
because of its analogy to the Dirac--Coulomb Hamiltonian,
which forms the basis for the description of
the relativistic hydrogen atom.
The latter Hamiltonian reads as 
\begin{equation}
\label{HDC}
H_{\rm DC} = \vec \alpha \cdot \vec p + \beta m_I
- \frac{Z \, \alpha}{r} \,,
\end{equation}
where $Z$ is the nuclear charge, and $\alpha$
is the fine-structure constant~\cite{ItZu1980}.
One immediate question
which comes to mind is why one cannot
simply insert the static gravitational potential
into the Dirac Hamiltonian, in analogy to Eq.~\eqref{Hseriously}, 
as one would do for the Coulomb potential 
in Eq.~\eqref{HDC}.
The answer is, in short, that the
Coulomb potential insertion can be traced to the
$U(1)$ covariant derivative,
in the sense of the replacement
$\ii \partial_\mu \to 
\ii \partial_\mu  - e \, A_\mu$
[see Eqs.~\eqref{cov1} and~\eqref{cov2}],
where $A_\mu$ is the external four-vector
potential, but gravity is not a $U(1)$
gauge theory like quantum electrodynamics (QED).

In order to put things into perspective,
we note that the Coulomb potential makes its way into the Dirac Hamiltonian 
by way of the covariant derivative,
which amounts to a replacement 
of the partial derivative by the $U(1)$ covariant derivative
\begin{equation}
\label{cov1}
\ii \partial_\mu \to \ii D_\mu =
\ii \partial_\mu  - e \, A_\mu \,.
\end{equation}
For $\mu = 0$ (timelike component), this implies that
\begin{equation}
\label{cov2}
\ii \partial_t \to 
\ii \partial_t - e \, \left(\frac{(-Z e)}{4 \pi r} \right) \,,
\end{equation}
where $Z$ is the nuclear charge number,
and $e$ is the electron charge ($e^2 = 4 \pi \alpha$
in natural units).

By contrast, note that
the gravitational interaction is not based on a $U(1)$ gauge
theory. A gauge transformation under gravitational coupling 
of a Dirac particle ensures the covariance with respect
to local Lorentz transformations [gauge group $SO(1,3)$], 
not $U(1)$ gauge 
transformations~\cite{Te1928,FoIw1928,Fo1929,Fo1929crasp,%
We1929,BrWh1957}.
For absolute clarity, we should remark that
a Foldy--Wouthuysen transformation
of the electrostatically coupled Hamiltonian~\eqref{HDC} leads to 
two Hamiltonians,
one for the particle, the other, for the antiparticle.
These describe the behavior 
of the electron and positron, respectively,
in a central binding Coulomb field.
They constitute special cases of the 
more general Eq.~(5) of Ref.~\citen{Je2018geonium}.
For the central-field gravitational problem, the starting point 
of the corresponding investigation has to be the Hamiltonian
given in Eq.~\eqref{HG}.

%
%
\section{Limitations of Einstein's Equivalence Principle}
\label{appb}

\setcounter{equation}{0}

After our intensive investigations of the relation of the 
equivalence principle to particle$\to$antiparticle
transformations, one might ask about 
further possible limitations to the general 
validity of Einstein's Equivalence Principle.
In view of the relations derived 
in Sec.~\ref{sec32} (particle-antiparticle symmetry), 
conceivable limitations would 
equally affect both particles and antiparticles.
Roughly speaking, one might ask if a 
fully deterministic theory, like general relativity, could
be fully compatible with a non-deterministic theory,
like quantum mechanics, given that 
the latter has to accommodate Heisenberg's uncertainty principle.

As already mentioned, but recalled for
convenience, the EEP states that {\em (i)}
the outcome of any local non-gravitational experiment is independent
of the velocity of the freely-falling reference frame in which it is performed
(local Lorentz invariance, LLI), and that, {\em (ii)},
the outcome of any local non-gravitational experiment is
independent of where and when in the universe it is performed
(local position invariance, LPI).
The question, though, is whether or not we precisely
know where and when in the Universe the experiment actually
was performed, given the Heisenberg uncertainty principle.
In particular, the nonvanishing positional
uncertainty of the electron wave packet in an atom
leads to an uncertainty in the exact position where, say,
a spectroscopically measured quantum jump took place.

Roughly speaking, one can say that the validity of 
LPI is limited by the fact that, due to 
quantum mechanics, at some level of accuracy, 
one cannot separate a non-gravitational
experiment from a gravitational one.
Let us illustrate this statement by way of example,
following Ref.~\citen{Je2018geonium}.
Relativistic geodesy~\cite{MaMu2013} is based on the comparison of the 
proper times $\dd \tau_1$ and $\dd \tau_2$ of two atomic 
clocks located at gravitational potentials $\Phi_1$ and $\Phi_2$, 
\begin{equation}
\label{EEP}
\frac{\dd \tau_1}{\dd \tau_2} = 
\frac{\sqrt{1 + 2 \Phi_1} }{\sqrt{1 + 2 \Phi_2} } \,.
\end{equation}
If the proper times measured by the atomic clocks 
follow the above relation, then full compatibility
with the EEP, notably, LPI, is achieved. 

According to Ref.~\citen{Je2018geonium}, the deviations can be expressed as 
follows,
\begin{equation}
\label{corr_term}
\frac{\dd \tau_1}{\dd \tau_2} = 
\frac{\displaystyle \sqrt{1 + 2 \Phi_1} + 
\left| \frac{\Phi_1}{\Phi_0} \right|^n \, C_n(M)}%
{\displaystyle 
\sqrt{1 + 2 \Phi_2} + \left| \frac{\Phi_2}{\Phi_0} \right|^n \, C_n(M)} \,.
\end{equation}
Here, $n$ refers to the power of the gravitational potential
at which the respective correction term
enters (no summation over $n$!).
In typical cases, one has either $n=2$ or $n=3$.
The coefficients $C_n(M)$ depend on the mass of the 
gravitational center, and on the effect under 
study (power law coefficient $n$).
The reference potential $\Phi_0$ is defined in
Eq.~\eqref{Phi0} below.
Some of the correction terms 
of the functional form 
$\left| \Phi_1/\Phi_0 \right|^n \, C_n(M)$
are due to the fact that the quantum mechanical 
wave function involved in the measurement
of the proper time intervals 
$\dd \tau_1$ and $\dd \tau_2$ ``wiggles'',
so that it is no longer possible to pinpoint
the exact location in the gravitational field 
where the measurement of the proper time interval 
took place.
Put differently, the Heisenberg uncertainty principle 
implies that at some point, one
cannot separate a gravitational from an
electromagnetic experiment. This leads to 
deviations from the ``perfect'' scaling implied
by Eq.~\eqref{EEP}.

We note that in Eqs.~(74), (76) and (77)
of Ref.~\citen{Je2018geonium},
one should understand the factor 
$| \Phi |^n$ as $|\Phi/\Phi_0|^n$;
the reference potential $\Phi_0$ used
in Eq.~\eqref{corr_term} is 
\begin{equation}
\label{Phi0}
\Phi_0 = \frac{G M_\oplus}{R_\oplus} \,,
\end{equation}
which is equal to the modulus of the 
gravitational potential on the Earth's surface.
The corrections which lead to a deviation 
from the scaling~\eqref{EEP}, according to 
Eq.~\eqref{corr_term}, have been analyzed 
in detail as $\delta E^{(i) \dots (iv)}$
in Ref.~\citen{Je2018geonium}.
A brief synopsis can be given as follows,
\begin{itemize}
\item
$\delta E^{(i)}$ is the quadrupole term,
evaluated for an excited atomic state
(it leads to an effect which scales as $|\Phi|^{n=3}$).
\item
$\delta E^{(ii)}$ is a second-order effect
due to the variation of the gravitational 
potential on distance scales commensurate with 
the size of the atom
(it leads to an effect with $n=2$).
\item
$\delta E^{(iii)}$ is the so-called 
Fokker--Planck correction, which is caused
by the coupling of the electron spin
to the gravitational field
(the power-law dependence has $n=3$).
\item
and, finally, $\delta E^{(iv)}$ is a 
first-order correction, due to the 
variation of the gravitational potential
on a distance scale commensurate with 
the molecular wave function;
it is nonvanishing only for oriented,
diatomic molecules (the effect has $n=2$).
\end{itemize}
We note that the corrections $\delta E^{(ii)}$ and
$\delta E^{(iv)}$ depend on the quantum-mechanical
positional uncertainty in the system and would thus 
vanish were it not for Heisenberg's uncertainty principle.
Of these, as shown in Table~1 of Ref.~\citen{Je2018geonium},
$\delta E^{(iv)}$ leads to 
coefficients in the range
\begin{equation}
C^{(iv)}_2(M_\oplus) \sim 10^{-20} \dots 10^{-18} 
\quad
\mbox{(oriented molecules)}
\end{equation}
for $\delta E^{(iv)}$ (for the Earth's gravitational
field). This effect could thus be measurable in the 
foreseeable future, as spectroscopic techniques
approach the $10^{-19}$ precision level~\cite{PrEtAl2013}.

Corrections $\delta E^{(i)}$, $\delta E^{(ii)}$ and
$\delta E^{(iv)}$, vanish in the hypothetical 
limit of a vanishing Bohr radius of the atom.
One might ask if full compatibility with the 
EEP could be restored if we could 
hypothetically ``switch off the Heisenberg principle''.
However, the correction $\delta E^{(iii)}$
comes into play, as a manifestation of the 
Fokker precession (FP) Hamiltonian, which, for an
electron interacting with the 
gravitational field of the Earth, reads as
\begin{equation}
\label{HFP}
H_{\rm FP} =
\frac{3 G M_\oplus}{4 m_e} \, \frac{\vec\sigma \cdot \vec L}{R^3} \,.
\end{equation}
This Hamiltonian describes the coupling of the 
electron spin to the gravitational field of the Earth. 
It is interesting to note that it is proportional 
to the inverse of the mass of the electron.
Let us suppose that we drive the 
$2P_{1/2} \to 2P_{3/2}$ spin-flip 
transition in atomic hydrogen with a laser,
an experiment which, {\em a priori},
would be understood as a fundamentally 
non-gravitational experiment.
The Fokker precession Hamiltonian,
which is the gravitational analogue of the 
Russell--Saunders spin-orbit coupling,
constitutes a gravitational coupling term
which cannot be ``switched off'' in nature,
not even in the limit of a vanishing Bohr 
radius. Its presence illustrates the 
statement made above, which implies that 
at some level of precision, it might be {\em in principle}
impossible to perform purely non-gravitational
experiments, because all particles 
involved in the experiments will also 
be subject to other fundamental forces. 
Numerical estimates lead to the 
result that 
\begin{equation}
C^{(iii)}_3(M_\oplus) \sim 10^{-44}
\quad
\mbox{(atoms and molecules)} \,.
\end{equation}
As such, the effect, while of utmost theoretical 
interest, will probably elude detection 
on atomic systems in the foreseeable future.
The theoretical interest is enhanced by the fact that 
the Fokker-precession term is generated by 
non-commutativity of momentum operators and the
gravitational potential, as a close inspection
of its derivation~\cite{JeNo2013pra} shows;
i.e., it is a true quantum effect 
beyond the scaling of the proper time
of the atomic clocks with $\sqrt{1 + 2 \Phi}$,
which can be obtained if we ignore the 
quantum commutators~\cite{Je2018geonium}.

A further remark is in order. Here, as well as in 
Ref.~\citen{Je2018geonium}, we have
concentrated on effects which persist even at zero temperature.
The first indication of a possible violation of the 
equivalence principle due to quantum effects
was in fact mentioned
in a series of papers~\cite{DoHoRo1984,DoHoRo1986,DoHo1987}
(see also Ref.~\citen{BlCaLaPe2019}), where the authors
analyzed a possible
violation of the Einstein equivalence principle
at finite temperature, for an electron in contact
with a heat bath of photons.
The calculations reported in~\cite{DoHoRo1984,DoHoRo1986,DoHo1987,BlCaLaPe2019} are
manifestly based on finite-temperature field
theory; all effects considered in 
Refs.~\citen{DoHoRo1984,DoHoRo1986,DoHo1987,BlCaLaPe2019}
vanish in the zero-temperature limit considered 
in the current work. 

Specifically, in 
Refs.~\citen{DoHoRo1984,DoHoRo1986,DoHo1987,BlCaLaPe2019},
the gravitational mass
of the particle at finite temperature is
derived based on the $\mu = \nu = 0$ component of the
energy-momentum tensor $\tau^{\mu\nu}$, and the coupling to the
gravitational field is described as in Eq.~(16) of 
Ref.~\citen{BlCaLaPe2019},
being proportional to a term of the form $h_{\mu\nu} \tau^{\mu\nu}$
in an equation of the form (in our notation),
\begin{equation}
\left( \ii \gamma^\mu \partial_\mu - m_I \right) \psi = 
\frac12 \, h_{\mu\nu} \tau^{\mu\nu}  \psi \,,
\end{equation}
where we assume vanishing temperature (hence, the 
vector $I_\mu$ in the notation of 
Ref.~\citen{BlCaLaPe2019} vanishes),
and $h_{\mu\nu}$ is taken as
$h_{\mu\nu} = 2 \; \Phi \; {\rm diag}(1,1,1,1)$ according to 
the text following Eq.~(16) of 
Ref.~\citen{BlCaLaPe2019},
where $\Phi$ is the gravitational potential.
It would be interesting to analyze if
this formalism is equivalent to the
covariant coupling discussed here in Sec.~\ref{sec2}.
Moreover, it would also be interesting to
verify if the effects described in
Refs.~\citen{DoHoRo1984,DoHoRo1986,DoHo1987,BlCaLaPe2019}
can be rederived, e.g., for a central gravitational
field, based on the approach described in Sec.~\ref{sec31}.
Note that the leading gravitational
coupling term written in Sec.~\ref{sec31}
is due to a simple mechanism which 
avoids the covariant derivative. Namely, it
comes from the ${\overline \gamma}^0$
which is proportional to 
$1/\sqrt{1 + 2 \Phi} \approx 1/(1 + \Phi)$
[see Eq.~\eqref{gravprep}].
The factor $1 + \Phi$
then meanders into the numerator of the right-hand side of the
Dirac equation, after solving for the time derivative
operator, and multiplies the (entire) mass term,
thus establishing the gravitational coupling in the 
central field [see Eq.~\eqref{gravdirect} and pertinent 
remarks following the mentioned equation].
It would be extremely interesting to
analyze, in detail, the relation of the conjectured 
temperature-dependent violation of the
Einstein equivalence principle 
(see Refs.~\citen{DoHoRo1984,DoHoRo1986,DoHo1987,BlCaLaPe2019})
to the formalism
of the gravitationally coupled Dirac equation,
laid out in the current work.
While further steps in this direction are beyond the scope of the current
investigation, we contend ourselves with the notion
that violations of the Einstein equivalence
principle due to quantum effects
(at finite temperature) have been discussed in the literature before.

%
%
\section{Relation of Our Considerations to the Penrose Conjecture}
\label{appc}

\setcounter{equation}{0}

Another potential limitation to the 
applicability of EEP comes from the Penrose
conjecture~\cite{Pe1996penrose,Pe1998penrose,Pe2014penrose}.
Roughly speaking, this conjecture deals with the 
following problem.
Due to the Heisenberg principle, the 
precise location of a particle described by a
quantum mechanical wave function is in principle 
endowed with uncertainty.
That means that we have a physical situation where
the exact shape of a mass distribution that needs to 
enter the Einstein equations is unknown (due to 
quantum uncertainty) unless the wave function has collapsed.
Yet, at some point, we need to know where the particle
is, or else we could not determine the space-time curvature 
(metric) around the objects.
In order to ensure that the uncertainty in determining the 
metric remains does not grow without bound, one postulates that 
the wave function must collapse at some point, in a non-constant
gravitational field.
This observation is the origin of the Penrose 
conjecture~\cite{Pe1996penrose,Pe1998penrose,Pe2014penrose}.

The conjecture then implies that collapse of the 
wave function should occur on a time scale
\begin{equation}
\label{tC}
t_C \sim \frac{\hbar}{E_G} \,,
\end{equation}
where $E_G$ is a measure of the 
gravitational energy contained in the 
un-collapsed wave function.
Various forms of $E_G$
have been discussed in the 
literature~\cite{Di1987,Di1989,Pe1996penrose,Pe1998penrose,Pe2014penrose,%
Je2018geonium}. An (unnumbered)
equation on p.~595 of Ref.~\cite{Pe1996penrose} puts
\begin{equation}
\label{EG1}
E_G = - G \, \int \dd^3 x \int \dd^3 y \,
\frac{\left[ \rho(\vec x) - \rho'(\vec x) \right] \,
\left[ \rho(\vec y) - \rho'(\vec y) \right]}%
{| \vec x - \vec y |} \,,
\end{equation}
where $\rho(\vec r)$ and $\rho'(\vec r)$ are the
two mass distributions, which represent 
possible outcomes of measurements of the position
of the particle, after the wave function collapses.
Equation~\eqref{EG1} describes the gravitational 
self-energy of the difference between the
two mass distributions.

Let us perform some order-of-magnitude estimates,
writing
\begin{subequations}
\begin{align}
\label{EG}
E_G =& \; -E_S - E'_S + E_I \,,
\\[0.1133ex]
\label{Sint}
E_S =& \; G \, \int \dd^3 x \int \dd^3 y \,
\frac{\rho(\vec x) \, \rho(\vec y) }{| \vec x - \vec y |} \,,
\\[0.1133ex]
\label{Sprime_int}
E_S' =& \; G \, \int \dd^3 x \int \dd^3 y \,
\frac{\rho'(\vec x) \, \rho'(\vec y) }{| \vec x - \vec y |} \,,
\\[0.1133ex]
\label{I_int}
E_I =& \; 2 G \, 
\int \dd^3 x \int \dd^3 y \,
\frac{\rho(\vec x) \, 
\rho'(\vec y) }{| \vec x - \vec y |} \,.
\end{align}
\end{subequations}
Here, $E_I$ has the interpretation of a gravitational
interaction energy between the two mass distributions,
while $E_S$ and $E'_S$ are the gravitational self-energies.

Order-of-magnitude estimates for typical quantum objects
can be given as follows:
\begin{itemize}
\item The mass $\int \dd^3 x \rho(\vec x) \sim m$
is of the order of the mass of an atomic 
nucleus, or, the proton mass of $\sim 10^{-27} \, {\rm kg}$.
\item The typical distance $ | \vec x - \vec y | $ in the 
self-energy integrals cannot
be smaller than a de Broglie wavelength the wave packet, or,
the size of an atom, which is the Bohr radius
$\sim (10^{-11}\dots10^{-10}) \, {\rm m}$.
\item The typical distance $ | \vec x - \vec y | $ in the 
interaction integrals cannot be smaller than the 
dimensions of technical device with which the 
atoms are being controlled, i.e., 
not smaller than a few nanometers, which 
is commensurate with today's microprocessor 
manufacturing standards,
$\sim 10^{-9} \, {\rm m}$.
\end{itemize}

Hence, for typical quantum systems, all three
entries in Eq.~\eqref{EG} are of the order of, 
or smaller, than
\begin{equation}
\label{EGestimate}
E_G \sim \frac{ 10^{-11} \times (10^{-27})^2 }{ 10^{-11} } =
10^{54} \, {\rm J}\,.
\end{equation}
For more detailed calculation based on the 
parameters of the Colella--Overhauser--Werner
experiment~\cite{OvCo1974,CoOvWe1975,BoWr1983,BoWr1984},
see Ref.~\citen{Je2018geonium}.
In the experiments~\cite{OvCo1974,CoOvWe1975,BoWr1983,BoWr1984}
a quantum wave packet is split in an interferometer on macroscopic distance scales
in a gravitational field
(for recently enhanced versions, which rely 
on atomic rather than neutron interferometry,
see Refs.~\cite{AsEtAl2017,OvEtAl2018}).
The mass distributions $\rho$ and $\rho'$ 
correspond to the two arms of the interferometer.
An important observation, compatible with the 
considerations above, is that 
the distance scale $| \vec x - \vec y|$, which enters
the self-energy integrals $E_S$ and $E_{S'}$,
is typically smaller than those which enter the the interaction 
integrals. Hence, the expression for $E_G$ 
given in Eq.~\eqref{EG} comes out as negative 
for typical configurations; one might have to eliminate 
the minus sign in Eq.~\eqref{EG1} or 
consider the modulus of the given quantity instead,
in order to obtain a positive value for the 
collapse time $t_C$.

On account of the smallness of the reduced
Planck constant, $\hbar \sim 10^{-34} \, {\rm Js}$, 
we have for typical quantum systems,
in view of Eq.~\eqref{EGestimate},
\begin{equation}
\label{tCestimate}
t_C \sim 10^{20} \, {\rm s} \,,
\end{equation}
which is longer than the age of the Universe.

These estimates also imply that gravitationally
induced wave collapse, according to the 
Penrose conjecture, does not lead to 
limitations for the functionality 
of quantum computers: Even if we control
on the order of $n = 10^{10}$~atoms coherently
on microscopic dimension, the estimate
given in Eq.~\eqref{tCestimate} would be reduced
by a factor $1/n^2 \sim 10^{-20}$, and still
could not ``collapse'' the wave function
on a time scale less than a second.


\begin{thebibliography}{10}

\bibitem{ALPHA}
ALPHA Collaboration {\em (Antihydrogen Laser Physics Apparatus)}, see the URL
  http://alpha-new.web.cern.ch.

\bibitem{AmEtAl2011}
\relax{C. Amole {\em et al.} [ALPHA Collaboration]}, {\em \relax{Resonant
  Quantum Transitions in Trapped Antihydrogen Atoms}},  Nature (London) {\bf
  483},  439--443  (2012).

\bibitem{AmEtAl2013}
\relax{C. Amole {\em et al.} [ALPHA Collaboration]}, {\em \relax{Description
  and first application of a new technique to measure the gravitational mass of
  antihydrogen}},  Nat. Commun. {\bf 4},  1785  (2013).

\bibitem{ATHENA}
ATHENA Collaboration {\em (ATHENA Antihydrogen Apparatus)}, see the URL
  http://athena.web.cern.ch/.

\bibitem{ATRAP}
ATRAP Collaboration {\em (Antihydrogen trap)}, see the URL
  http://home.cern/science/\allowbreak{}experiments/atrap.

\bibitem{GaEtAl2008}
\relax{G. Gabrielse {\em et al.} [ATRAP Collaboration]}, {\em
  \relax{Antihydrogen Production within a Penning-Ioffe Trap}},  Phys. Rev.
  Lett. {\bf 100},  113001  (2008).

\bibitem{Ke2008}
\relax{A. Kellerbauer {\em et al.} [AEGIS Proto-Collaboration]}, {\em
  \relax{Proposed antimatter gravity measurement with an antihydrogen beam}},
  Nucl. Instrum. Methods Phys. Res. B {\bf 266},  351--356  (2008).

\bibitem{AGELOIshort}
\relax{A. D. Cronin {\em et al.} [AGE Collaboration]}, Letter of Intent:
  Antimatter Gravity Experiment (AGE) at Fermilab (2009).

\bibitem{AGELOI}
\relax{A. D. Cronin {\em et al.} [AGE Collaboration]}, Letter of Intent:
  Antimatter Gravity Experiment (AGE) at Fermilab (2009), available at the URL
  http://www.fnal.gov/\allowbreak{}directorate/\allowbreak{}program\_planning/%
  \allowbreak{}Mar2009PACPublic/\allowbreak{}AGELOIFeb2009.pdf; see also the
  URL
  http://www.phy.duke.edu/\~{}phillips/\allowbreak{}gravity/\allowbreak{}frameIndex.html.

\bibitem{MoPlSo1998}
P.~J. Mohr, G. Plunien, and G. Soff, {\em \relax{QED corrections in heavy
  atoms}},  Phys. Rep. {\bf 293},  227--372  (1998).

\bibitem{Di1928a}
P.~A.~M. Dirac, {\em \relax{The Quantum Theory of the Electron}},  Proc. Roy.
  Soc. London, Ser. A {\bf 117},  610--624  (1928).

\bibitem{Di1928b}
P.~A.~M. Dirac, {\em \relax{The Quantum Theory of the Electron}},  Proc. Roy.
  Soc. London, Ser. A {\bf 118},  351--361  (1928).

\bibitem{ItZu1980}
C. Itzykson and J.~B. Zuber, {\em \relax{Quantum Field Theory}} (McGraw-Hill,
  New York, 1980).

\bibitem{An1933}
C.~D. Anderson, {\em \relax{The Positive Electron}},  Phys. Rev. {\bf 43},
  491--409  (1933).

\bibitem{Ma2016}
J. Maruani, {\em \relax{The Dirac Electron: From Quantum Chemistry to Holistic
  Cosmology}},  J. Chin. Chem. Soc. {\bf 63},  33--48  (2016).

\bibitem{Ko1996}
M. Kowitt, {\em \relax{Gravitational Repulsion and Dirac Antimatter}},  Int. J.
  Theor. Phys. {\bf 35},  605--631  (1996).

\bibitem{FoWu1950}
L.~L. Foldy and S.~A. Wouthuysen, {\em \relax{On the Dirac Theory of Spin $1/2$
  Particles and Its Non-Relativistic Limit}},  Phys. Rev. {\bf 78},  29--36
  (1950).

\bibitem{JeWu2012epjc}
U.~D. Jentschura and B.~J. Wundt, {\em \relax{Localizability of Tachyonic
  Particles and Neutrinoless Double Beta Decay}},  Eur. Phys. J. C {\bf 72},
  1894  (2012) [13 pages].

\bibitem{Be1947}
H.~A. Bethe, {\em \relax{The Electromagnetic Shift of Energy Levels}},  Phys.
  Rev. {\bf 72},  339--341  (1947).

\bibitem{Je2013}
U.~D. Jentschura, {\em \relax{Gravitationally Coupled Dirac Equation for
  Antimatter}},  Phys. Rev. A {\bf 87},  032101  (2013), [Erratum Phys.~Rev.~A
  {\bf 87}, 069903(E) (2013)].

\bibitem{JeNo2013pra}
U.~D. Jentschura and J.~H. Noble, {\em \relax{Nonrelativistic Limit of the
  Dirac--Schwarzschild Hamiltonian: Gravitational Zitterbewegung and
  Gravitational Spin--Orbit Coupling}},  Phys. Rev. A {\bf 88},  022121
  (2013).

\bibitem{JeNo2014jpa}
U.~D. Jentschura and J.~H. Noble, {\em \relax{Foldy--Wouthuysen transformation,
  scalar potentials and gravity}},  J. Phys. A {\bf 47},  045402  (2014).

\bibitem{NoJe2015tach}
J.~H. Noble and U.~D. Jentschura, {\em \relax{Ultrarelativistic Decoupling
  Transformation for Generalized Dirac Equations}},  Phys. Rev. A {\bf 92},
  012101  (2015).

\bibitem{NoJe2016}
J.~H. Noble and U.~D. Jentschura, {\em \relax{Dirac Hamiltonian and
  Reissner--Nordstr\"{o}m Metric: Coulomb Interaction in Curved Space--Time}},
  Phys. Rev. A {\bf 93},  032108  (2016).

\bibitem{Je2018geonium}
U.~D. Jentschura, {\em \relax{Gravitational Effects in $g$ Factor Measurements
  and High--Precision Spectroscopy: Limits of Einstein's Equivalence
  Principle}},  Phys. Rev. A {\bf 98},  032508  (2018).

\bibitem{Te1928}
H. Tetrode, {\em \relax{Allgemein-relativistische Quantentheorie des
  Elektrons}},  Z. Phys. {\bf 50},  336--346  (1928).

\bibitem{FoIw1928}
V. Fock and D. Iwanenko, {\em \relax{\"{U}ber eine m\"{o}gliche geometrische
  Deutung der relativistischen Quantentheorie}},  Z. Phys. {\bf 56},  798--802
  (1929).

\bibitem{Fo1929}
V. Fock, {\em \relax{Geometrisierung der Diracschen Theorie des Elektrons}},
  Z. Phys. {\bf 57},  261--277  (1929).

\bibitem{Fo1929crasp}
V. Fock and D. Ivanenko, {\em \relax{G\'{e}om\'{e}trie quantique lin\'{e}aire
  to d\'{e}placement parall\`{e}le}},  C. R. Acad. Sci. Paris {\bf 188},
  1470--1472  (1929).

\bibitem{We1929}
H. Weyl, {\em \relax{Gravitation and the Electron}},  Proc. Natl. Acad. Sci.
  USA {\bf 15},  323--334  (1929).

\bibitem{BrWh1957}
D.~R. Brill and J.~A. Wheeler, {\em \relax{Interaction of Neutrinos and
  Gravitational Fields}},  Rev. Mod. Phys. {\bf 29},  465--479  (1957).

\bibitem{Bl2018}
A.~S. Blum,  in {\em \relax{Quantum Gravity in the First Half of the Twentieth
  Century}}, edited by A.~S. Blum and D. Rickles (Edition Open Sources,
  Max--Planck--Institute for the History of Science, 2018), pp.\ 49--56.

\bibitem{Bo2011}
M. Bojowald, {\em \relax{Canonical Gravity and Applications}} (Cambridge
  University Press, Cambridge, 2011).

\bibitem{Iv1969b}
O.~S. Ivanitskaya, {\em \relax{Lorentzian basis and gravitational effects in
  Einstein’s theory of gravity (in Russian)}} (Nauka i Technika, Minsk, USSR,
  1969).

\bibitem{IvMiVl1985}
O.~S. Ivanitskaya, N.~V. Mitskievic, and Y.~S. Vladimirov,  in {\em
  \relax{Proceedings of the 114th Symposium of the International Astronomical
  Union held in Leningrad, USSR, May 1985}}, edited by J. Kovelevsky and V.~A.
  Brumberg (Kluwer, Dordrecht, 1985), pp.\ 177--186.

\bibitem{GaEtAl1999}
G. Gabrielse, A. Khabbaz, D.~S. Hall, C. Heimann, H. Kalinowsky, and W. Jhe,
  {\em \relax{Precision Mass Spectroscopy of the Antiproton and Proton Using
  Simultaneously Trapped Particles}},  Phys. Rev. Lett. {\bf 82},  3198--3201
  (1999).

\bibitem{UlEtAl2014}
S. Ulmer, C. Smorra, A. Mooser, K. Franke, H. Nagaharma, G. Schneider, T.
  Higuchi, S. Van~Grop, K. Blaum, Y. Matsuda, W. Quint, J. Walz, and Y.
  Yamazaki, {\em \relax{High-precision comparison of the antiproton-to-proton
  charge-to-mass ratio}},  Nature (London) {\bf 524},  196--199  (2015).

\bibitem{Je2011radii}
U.~D. Jentschura, {\em \relax{Proton Radius, Darwin-Foldy Term and Radiative
  Corrections}},  Eur. Phys. J. D {\bf 61},  7--14  (2011).

\bibitem{BjDr1964}
J.~D. Bjorken and S.~D. Drell, {\em \relax{Relativistic Quantum Mechanics}}
  (McGraw-Hill, New York, 1964).

\bibitem{BjDr1965}
J.~D. Bjorken and S.~D. Drell, {\em \relax{Relativistic Quantum Fields}}
  (McGraw-Hill, New York, 1965).

\bibitem{JeAd2019}
U.~D. Jentschura and G.~S. Adkins, {\em \relax{Quantum electrodynamics and
  beyond: Relativity, atoms, lasers and gravity}} (World Scientific, Singapore,
  scheduled for 2019).

\bibitem{Wi1974prd}
C.~M. Will, {\em \relax{Gravitational red-shift measurements as tests of
  nonmetric theories of gravity}},  Phys. Rev. D {\bf 10},  2330--2337  (1974).

\bibitem{Go1963}
H. Goldstein, {\em \relax{Klassische Mechanik}} (Akademische
  Verlagsgesellschaft, Frankfurt am Main, 1963).

\bibitem{Je2017book}
U.~D. Jentschura, {\em \relax{Advanced Classical Electrodynamics: Green
  functions, regularizations, multipole decompositions}} (World Scientific,
  Singapore, 2017).

\bibitem{MaMu2013}
E. Mai and J. M\"{u}ller, {\em \relax{General Remarks on the Potential Use of
  Atomic Clocks in Relativistic Geodesy}},  ZFV---Zeitschrift f\relax{\"{u}}r
  Geod\relax{\"{a}}sie, Geoinformation und Landmanagement {\bf 138},  255--267
  (2013).

\bibitem{PrEtAl2013}
K. Predehl, G. Grosche, S.~F.~F. Raupach, S. Droste, O. Terra, J. Alnis, T.
  Legero, T.~W. H\"{a}nsch, T. Udem, R. Holzwarth, and H. Schnatz, {\em
  \relax{A 920-Kilometer Optical Fiber Link for Frequency Metrology at the 19th
  Decimal Place}},  Science {\bf 336},  441--444  (2013).

\bibitem{DoHoRo1984}
J.~F. Donoghue, B.~R. Holstein, and R.~W. Robinett, {\em \relax{Renormalization
  of the energy-momentum tensor and the validity of the equivalence principle
  at finite temperature}},  Phys. Rev. D {\bf 30},  2561--2572  (1984).

\bibitem{DoHoRo1986}
J.~F. Donoghue, B.~R. Holstein, and R.~W. Robinett, {\em \relax{Gravitational
  coupling at finite temperature}},  Phys. Rev. D {\bf 34},  1208--1209
  (1986).

\bibitem{DoHo1987}
J.~F. Donoghue and B.~R. Holstein, {\em \relax{Aristotle was right: heavier
  objects fall faster}},  Eur. J. Phys. {\bf 8},  105--113  (1987).

\bibitem{BlCaLaPe2019}
M. Blasone, S. Capozziello, G. Lambiase, and L. Petruzziello, {\em
  \relax{Equivalence principle violation at finite temperature in scalar-tensor
  gravity}},  Eur. Phys. J. Plus {\bf 134},  169  (2019).

\bibitem{Pe1996penrose}
R. Penrose, {\em \relax{On Gravity's Role in Quantum State Reduction}},  Gen.
  Relativ. Gravit. {\bf 28},  581--600  (1996).

\bibitem{Pe1998penrose}
R. Penrose, {\em \relax{Quantum computation, entanglement and state
  reduction}},  Phil. Trans. R. Soc. Lond. A {\bf 356},  1927--1939  (1998).

\bibitem{Pe2014penrose}
R. Penrose, {\em \relax{On the Gravitization of Quantum Mechanics 1: Quantum
  State Reduction}},  Found. Phys. {\bf 44},  557--575  (2014).

\bibitem{Di1987}
L. Di\'{o}si, {\em \relax{A universal master equation for the gravitational
  violation of quantum mechanics}},  Phys. Lett. A {\bf 120},  377--381
  (1987).

\bibitem{Di1989}
L. Di\'{o}si, {\em \relax{Models for universal reduction of macroscopic quantum
  flutuations}},  Phys. Rev. A {\bf 40},  1165--1174  (1989).

\bibitem{OvCo1974}
A.~W. Overhauser and R. Colella, {\em \relax{Experimental Test of
  Gravitationally Induced Quantum Interference}},  Phys. Rev. Lett. {\bf 33},
  1237--1239  (1974).

\bibitem{CoOvWe1975}
R. Colella, A.~W. Overhauser, and S.~A. Werner, {\em \relax{Observation of
  Gravitationally Induced Quantum Interference}},  Phys. Rev. Lett. {\bf 34},
  1472--1474  (1975).

\bibitem{BoWr1983}
U. Bonse and T. Wroblewski, {\em \relax{Measurement of Neutron Quantum
  Interference in Noninertial Frames}},  Phys. Rev. Lett. {\bf 51},  1401--1404
   (1983).

\bibitem{BoWr1984}
U. Bonse and T. Wroblewski, {\em \relax{Dynamical diffraction effects in
  noninertial neutron interferometry}},  Phys. Rev. D {\bf 30},  1214--1217
  (1984).

\bibitem{AsEtAl2017}
P. Asenbaum, C. Overstreet, T. Kovachy, D.~D. Brown, J.~M. Hogan, and M.~A.
  Kasevich, {\em \relax{Phase Shift in an Atom Interferometer due to Spacetime
  Curvature across its Wave Function}},  Phys. Rev. Lett. {\bf 118},  183602
  (2017).

\bibitem{OvEtAl2018}
C. Overstreet, P. Asenbaum, T. Kovachy, R. Notermans, J.~M. Hogan, and M.~A.
  Kasevich, {\em \relax{Effective Inertial Frame in an Atom Interferometric
  Test of the Equivalence Principle}},  Phys. Rev. Lett. {\bf 120},  183604
  (2018).

\end{thebibliography}
\end{document}